# Remarkable Daytime Sub-ambient Radiative Cooling in BaSO$_4$ Nanoparticle Films and Paints

*Xiangyu Li* [†,‡], *Joseph Peoples* [†,‡], *Peiyan Yao* [†,‡] *and Xiulin Ruan* [†,‡,\*]
[†] *School of Mechanical Engineering, Purdue University, West Lafayette, IN 47907, USA*
[‡] *Birck Nanotechnology Center, Purdue University, West Lafayette, IN 47907, USA*
[\*] *Corresponding Author, Email:* ruan@purdue.edu

## ABSTRACT

Radiative cooling is a passive cooling technology that offers great promises to reduce space cooling cost, combat the urban island effect and alleviate the global warming. To achieve passive daytime radiative cooling, current state-of-the-art solutions often utilize complicated multilayer structures or a reflective metal layer, limiting their applications in many fields. Attempts have been made to achieve passive daytime radiative cooling with single-layer paints, but they often require a thick coating or show partial daytime cooling. In this work, we experimentally demonstrate remarkable full daytime sub-ambient cooling performance with both BaSO$_4$ nanoparticle films and BaSO$_4$ nanocomposite paints. BaSO$_4$ has a high electron bandgap for low solar absorptance and phonon resonance at 9 μm for high sky window emissivity. With an appropriate particle size and a broad particle size distribution, BaSO$_4$ nanoparticle film reaches an ultra-high solar reflectance of 97.6% and high sky window emissivity of 0.96. During field tests, BaSO$_4$ film stays more than 4.5°C below ambient temperature or achieves average cooling power of 117 W/m$^2$. BaSO$_4$-acrylic paint is developed with 60% volume concentration to enhance the reliability in outdoor applications, achieving solar reflectance of 98.1% and sky window emissivity of 0.95. Field tests indicate similar cooling performance to the BaSO$_4$ films. Overall, our BaSO$_4$-acrylic paint shows standard figure of merit of 0.77 which is among the highest of radiative cooling solutions, while providing great reliability, the convenient paint form, ease of use and the compatibility with commercial paint fabrication process.

**TOC GRAPHICS**

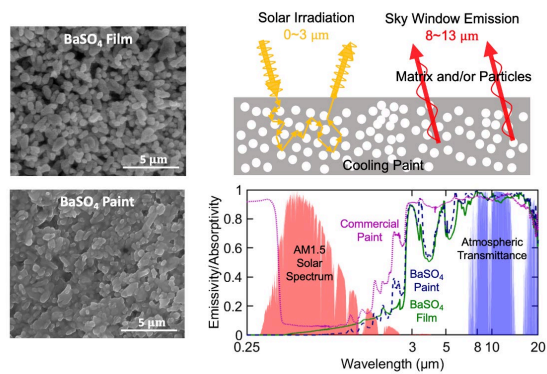

Radiative cooling has shown great promises to reduce the cost of space cooling in both residential and commercial applications.[1] Contrary to active cooling which requires electricity to drive a refrigeration cycle, radiative cooling utilizes the atmospheric transparent window (the "sky window") to emit thermal radiation directly to the deep sky without consuming any energy.[2] Its passive nature has the potential to reduce the urban island effect and alleviate the global warming. If the thermal emission through the sky window exceeds the solar absorption, the surface can maintain cooler than ambient even under direct sunlight. Early studies of radiative cooling paints lasted for decades, but none of the paints achieved full daytime radiative cooling.[2–11] Among these, one study showed 2°C below ambient cooling on a winter day with a thin layer of $TiO_2$ on aluminum substrate, but the high solar reflectance should primarily come from the metal substrate rather than the paint itself.[9] Many paints were based on $TiO_2$ with low particle concentration, and the radiative cooling performance was limited by insufficient solar reflection due to the solar absorptance in the ultra-violet (UV) band. For this reason, wide bandgap materials were explored as fillers [12,13] to eliminate the UV absorption, while their smaller refractive index makes photon scattering weaker. Heat reflective paints have also been developed, but their solar reflection is still limited below 91% and does not show full-daytime sub-ambient cooling.[11,14] Alternatively, photonic structures and multilayers have recently demonstrated full daytime sub-ambient cooling capability, which stimulated renewed interest in radiative cooling.[15,16] Other studies explored scalable non-paint approaches such as dual layers including a metal layer,[17–19] polyethylene aerogel [20] and delignified wood [21]. However, these approaches are limited in one or more aspects such as complicated structure, involvement of metallic layer, and large thickness, preventing them from many applications. In light of this, creating high-performance radiative cooling paints is still pertinent task. Recently, non-metal dual-layer designs were proposed to consist of top $TiO_2$ layer for solar reflectance and bottom layer for thermal emission, which achieved partial daytime cooling without metallic components.[22,23] A compact film of $SiO_2$ nanoparticles was fabricated as a single-layer coating with partial daytime cooling capability.[24] A polytetrafluoroethylene (PTFE) nanoparticle coating with a silver layer reached a record-high solar reflectance of 99%.[25] Paint-like porous polymers were developed with full-daytime cooling.[26] A strategy was proposed to further enhance the solar reflectance in particle-matrix paint by adopting a broad particle size distribution rather than one single size.[27] Combining a broad particle size distribution and a high filler concentration, $CaCO_3$-acrylic paint was developed and demonstrated full daytime sub-ambient cooling.[28–30] Another study also proposed high solar reflectance in wide bandgap nanoparticle paints with high filler concentrations.[31] Considering the existing studies, developing high-performance single-layer coatings that are thin, low cost, easy to apply, and scalable is still a challenging and urgent task to fully utilize radiative cooling in a wide range of applications.

In this work, we experimentally demonstrate full daytime sub-ambient cooling with $BaSO_4$ nanoparticle film and $BaSO_4$-acrylic paints. We choose $BaSO_4$ due to its high electron bandgap for low solar absorptance and phonon resonance at 9 μm for high sky window emissivity. By adopting an appropriate particle size and a broad particle size distribution, we achieve a high solar reflectance of 97.6% and a high sky window emissivity of 0.96 with $BaSO_4$ nanoparticle film. Field tests indicate surface temperature more than 4.5°C below ambient temperature or average cooling power of 117 $W/m^2$, among the highest cooling power reported. To enhance the reliability of the coating, $BaSO_4$-acrylic paint is developed with 60% volume concentration. The high filler concentration and the broad particle size distribution are added to compensate the low refractive index of $BaSO_4$ compared to $TiO_2$, leading to solar reflectance of 98.1% and sky window emissivity of 0.95. During field tests, the $BaSO_4$ paint yields similarly high cooling performance. Our $BaSO_4$-acrylic paint shows standard figure of merit of 0.77 which is among the highest of radiative cooling solutions, while providing great reliability, the convenient paint form, ease of use and the compatibility with commercial paint fabrication process. The results were also included in a provisional patent filed on October 3, 2018 and a non-provisional international patent application (PCT/US2019/054566) filed on October 3, 2019 and published on April 9, 2020.[29]

Commercial white paints such as $TiO_2$-acrylic paint failed to achieve full daytime cooling, which is attributed to its high solar absorption in the UV band (due to the 3.2 eV electron bandgap of TiO2) and near-infrared (NIR) band (due to acrylic absorption). In this work, we fabricated a BaSO4 particle film with a thickness of 150 μm on a silicon wafer (Figure 1a) along with a commercial white paint (DutchBoy Maxbond UltraWhite Exterior Acrylic Paint). An SEM image of the BaSO4 film is shown in Figure 1b, where air voids were introduced in the film. The interfaces between BaSO4 nanoparticles and air void enhance the photon scattering in the film, thus increase the overall solar reflectance. To improve the reliability of the coating under long-term outdoor exposure, a commercial paint form as the filler-matrix composite is often preferred. A key challenge to adopt BaSO4 as a filler material in polymer matrix is the low refractive index of BaSO4 compared to that of TiO2. To enable strong scattering in the composite, we adopted a particle volume concentration of 60%, which is considerably higher than those in commercial paints. Additionally, the broad particle size distribution contributes to the solar reflectance.[27] The BaSO4-acrylic nanocomposite paint is shown in Figure 1a with an SEM image in Figure 1c. The addition of acrylic helps bond the fillers and leads to a better reliability. Some air voids were present in the BaSO4 paint, which also increases the solar reflectance. The particle size distribution (398 ± 130 nm) of the BaSO4 particles was characterized with SEM images.

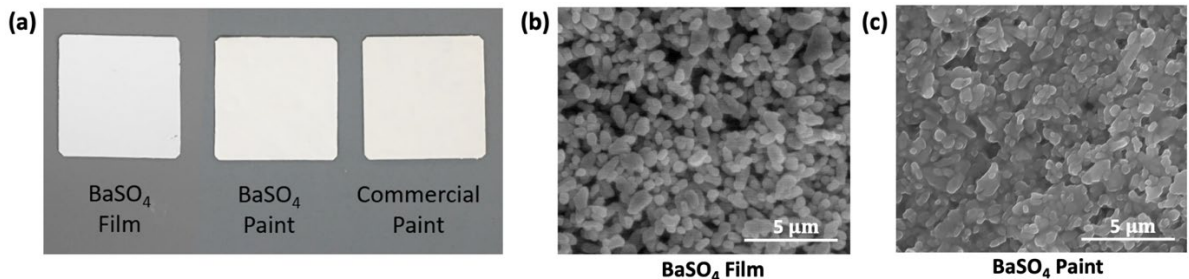

**Figure 1.** Radiative cooling coatings and SEM images. (a) BaSO4 film sample, BaSO4-acrylic paint sample and white commercial paint. The BaSO4 film is 150 μm thick on a silicon wafer. The BaSO4 paint and commercial paint are free-standing samples with a thickness of 400 μm. All samples are 5 cm squares. (b) An SEM image of the BaSO4 film sample. (c) An SEM image of the BaSO4-acrylic paint sample with 60% filler concentration. (b,c) The particle size distribution (398 ± 130 nm) was estimated based on the SEM images. Air voids were introduced in both the BaSO4 film and the BaSO4 paint.

To achieve full daytime sub-ambient cooling, both high solar reflectance (contributed by particles) and high sky window emissivity (contributed by particles and/or matrix) are essential as illustrated in Figure 2a. Here we adopted $BaSO_4$ with a high electron band gap ~6 eV to reduce the absorption in UV band. Due to a phonon resonance at 9 μm which is in the sky window, engineering the particle size can allow a single layer of $BaSO_4$ particle film to function both as a sky window emitter and a solar reflector. Therefore, the matrix is not needed to achieve daytime sub-ambient cooling. The lack of acrylic matrix also reduces the NIR absorption. The average particle size of $BaSO_4$ was chosen as 400 nm to reflect both visible and NIR range of the solar irradiation. A broad particle size distribution was adopted to further enhance the solar reflectance. Detailed theoretical and experimental studies can be found in previous studies.[27,28] Overall, the $BaSO_4$ film reached a solar reflectance of 97.6% and an emissivity of 0.96 in the sky window, as shown in Figure 2b. The solar reflectance is significantly higher than the commercial white paint (DutchBoy Maxbond UltraWhite Exterior Acrylic Paint, 400 μm thickness), especially in the UV and NIR range. Although the DutchBoy paint is not a heat reflective commercial paint, the solar reflectance of the $BaSO_4$ film is still substantially higher than those of the commercial heat reflective paints which show solar reflectance of around 80% to 91%.[11,14] The reflectance is also higher than that of the recently reported $CaCO_3$-acrylic paint.[28,32] The silicon substrate was intended only as a supporting substrate, not to increase the solar reflectance, nor to emit in the sky window. To avoid the substrate effect on the cooling performance, we characterized the thickness-dependent solar reflectance with different substrates, shown in Figure 2c. Night-time cooling performance was also characterized with different substrates in the Supplementary Note 1 and Figure S1. For the $BaSO_4$ paint, a standalone paint sample of 400 μm reached similar optical properties (98.1% solar reflectance, 0.95 sky window emissivity) with a high filler concentration of 60% and a broad size distribution. A Monte Carlo simulation with modified Lorentz-Mie theory was run to help illustrate the physics behind the strong solar reflectance of the paint, and the results are shown in Figure 2d.[27,28,33] With the same filler concentration, the simulation demonstrated a broader particle size distribution further improved the overall solar reflectance. The simulation slightly underestimated the solar reflectance as it cannot capture the effect of air voids. The coating thickness of the $BaSO_4$ paint sample was 400 μm, to ensure the optical properties were substrate independent. Thinner coatings were fabricated on transparent polyethylene terephthalate (PET) film using a film applicator to control the wet film thickness. With low transmittance of the paint films and similar refractive index between the paint and the PET film, the PET substrate had a negligible effect on the solar reflectance. With a thinner coating thickness of 200 μm, 224 μm and 280 μm, the solar reflectance reached 95.8%, 96.2% and 96.8%, respectively (Figure 2e). Monte Carlo simulation results showed a similar trend to the experimental data, both following the diffusion behavior of the transmission.[24,34] The simulation slightly underestimated the solar reflectance, likely because the air voids introduced in the paint were not captured by the Monte Carlo simulation. Future studies can aim to include such effect to predict the solar reflectance more accurately.

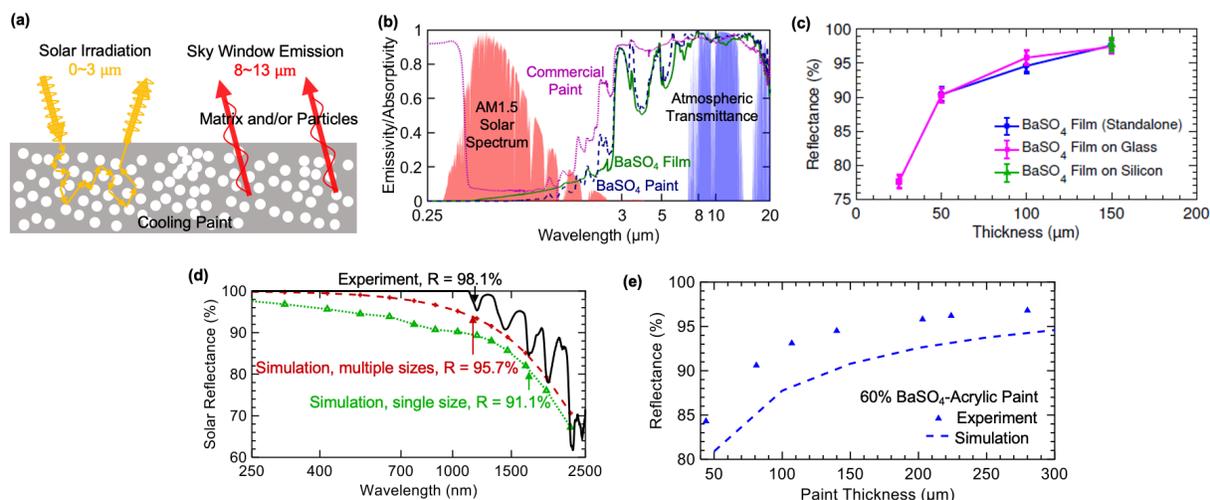

**Figure 2.** Radiative cooling schematic, spectral characterization results and Monte Carlo simulation on the solar reflectance. (a) To achieve a high cooling power with passive radiative cooling, both high solar reflectance and high sky window emissivity are needed. The solar reflectance is contributed by filler materials, while the sky window emissivity can come from fillers and/or matrix. For particle films, the particles have to both reflect solar light and emit in the sky window. (b) The emissivities of $BaSO_4$ film and $BaSO_4$ paint were characterized, compared with commercial white paint (DutchBoy Maxbond UltraWhite Exterior Acrylic Paint, thickness of 400 μm) from 0.25 μm to 20 μm. Both the particle film and nanocomposite paint showed significant enhancement of solar reflectance while maintaining high sky window emissivity. (c) The solar reflectances of the $BaSO_4$ films with different thicknesses and substrates were measured, demonstrating that the solar reflectance of the $BaSO_4$ film at 150 μm is substrate independent. (d) Monte Carlo simulation of the $BaSO_4$ paint with 400 μm thickness demonstrated that both high filler concentration and broad particle size distribution increased the overall solar reflectance. (e) The solar reflectances of the $BaSO_4$-acrylic paint with 60% particle concentration and different film thicknesses were compared with the Monte Carlo simulation results. The thin paint coatings are supported by PET films.

Onsite field tests were performed to demonstrate the full daytime sub-ambient cooling of the $BaSO_4$ film. In Figure 3a, the $BaSO_4$ film achieved full daytime cooling below the ambient temperature with a peak solar irradiation of 907 W/m² in West Lafayette, IN on March 14-16, 2018 with 42% humidity at noon. The temperature of the sample dropped 10.5°C below the ambient temperature during the nights, and stayed 4.5°C below ambient even at the peak solar irradiation, whereas the commercial paint rose 6.8°C above the ambient temperature. A direct measurement of the cooling power in Reno, NV on July 28, 2018 showed that the cooling power reached an average of 117 W/m² over a 24-hour period with 10% humidity at noon, shown in Figure 3b. We observed similar daytime cooling power to nighttime without solar irradiation, both above 110 W/m². Thermal emission power increases with higher surface temperature in the daytime, which compensates the higher solar absorption. Thus, simply reporting the cooling power without considering the surface temperature can be a misleading measure of the cooling performance. In this case, the thermal emissive power of the BaSO$_4$ film reaches 106 W/m$^2$ at 15°C. Overall, our BaSO$_4$ film can maintain a constant high cooling power regardless of the solar irradiation. We further demonstrated the cooling performance of BaSO$_4$ paint with onsite field tests, as shown in Figure 3c and Figure 3d. The BaSO$_4$ paint remained cooler than ambient for more than 24 hours under the peak solar irradiation of 993 W/m$^2$ (around 50% humidity at 12:00 PM). The cooling power measurement showed an average cooling power over 80 W/m$^2$ with the surface temperature as low as -10°C, equivalent to a cooling power of 113 W/m$^2$ at 15°C (see more details in the Supplementary Note 2).

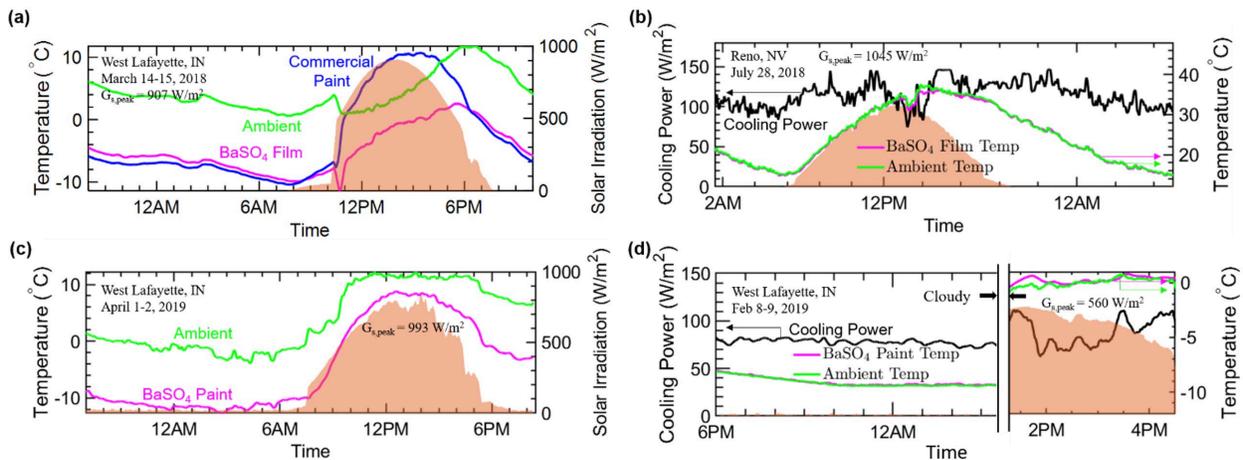

**Figure 3.** Field test results of the BaSO$_4$ nanoparticle film and BaSO$_4$-acrylic nanocomposite paint. (a) The temperatures of the BaSO$_4$ nanoparticle film and commercial white paint were compared to the ambient temperature for over 24 hours. (b) The cooling power was directly measured for the BaSO$_4$ nanoparticle film using a feedback heater. (c) The temperature of BaSO$_4$ paint was compared with the ambient temperature. (d) The cooling power of BaSO$_4$ paint was measured in both daytime and nighttime. The orange regions stand for the solar irradiation intensity.

The figure of merit RC was used here to fairly access the radiative cooling performance independent of weather conditions, as [28]

$$RC = \epsilon_{Sky} - r(1 - R_{Solar})$$

where $\epsilon_{Sky}$ is the sky window emissivity, $R_{Solar}$ is the solar reflectance, and $r$ is the ratio of the solar irradiation power over the blackbody emissive power through the sky window. With surface temperature of 300 K and $r$ set as 10, our BaSO$_4$ film and BaSO$_4$ paint reach the standard RC of 0.72 and 0.77, respectively, which are higher than state-of-the-art radiative cooling solutions as 0.32 [16], 0.53 [18], 0.35 [19], 0.49 [28], and 0.57 [35].

To demonstrate the reliability of our BaSO$_4$ paint, we conducted abrasion test, outdoor weathering and viscosity characterizations. The abrasion tests were performed according to ASTM D4060 with a Taber Abraser Research Model, and the results are shown in Figure 4a.[36] A pair of abrasive wheels were placed on the sample surface with 250 g load on each wheel. Mass loss was measured, and wheel refacing was done every 500 cycles. The wear index was defined as the weight loss (mg) for every 1000 cycles. The BaSO$_4$ paint reached a wear index of 150, comparable to the commercial exterior paint with wear index of 104. The weathering test was conducted by exposing the BaSO$_4$ paint outdoors for 3 weeks (Figure 4b). The solar reflectance remained the same within the experimental uncertainty. The sky window emissivity was measured to be 0.95 both at the beginning and the end of the testing period. Additionally, a running water test was included in the Video S1. In Figure 4c, we measured the viscosity of the BaSO$_4$ paint, which was similar to the commercial paints.[37] The Video S1 further demonstrated that the BaSO$_4$ paint was able to be brushed and dried similarly to the commercial paints. A detailed cost analysis of the BaSO$_4$ paint is included in the Supplementary Note 3.

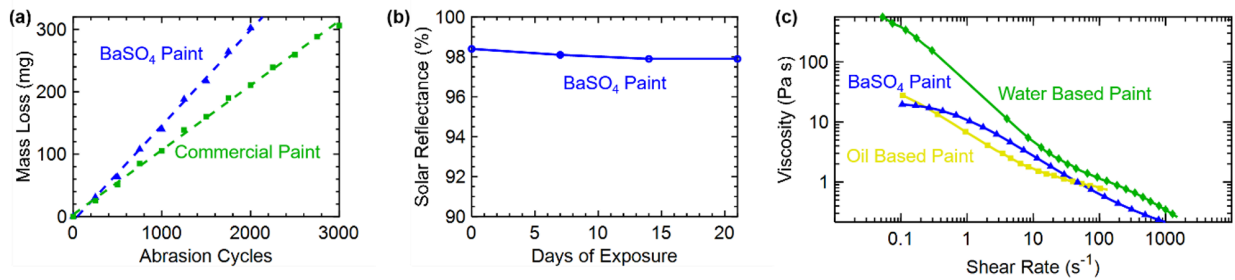

**Figure 4.** Reliability tests of the BaSO$_4$ paint. (a) Abrasion tests were conducted with BaSO$_4$ paint and commercial exterior paint according to ASTM D4060.[36] Our BaSO$_4$ paint demonstrated comparable abrasion resistance with commercial exterior paint. (b) We exposed the BaSO$_4$ paint outdoor for a 3-week period. (c) The viscosity of the BaSO$_4$ paint was characterized and compared with those of the commercial paints.[37]

In this work, we experimentally demonstrate full daytime radiative cooling with BaSO$_4$ nanoparticle film and BaSO$_4$-acrylic nanocomposite paint. By adopting an appropriate particle size and a broad particle size distribution, we achieve a high solar reflectance of 97.6% and high sky window emissivity of 0.96 with BaSO$_4$ nanoparticle film. Onsite field tests indicate surface temperature more than 4.5°C below ambient temperature or average cooling power of 117 W/m$^2$, among the highest cooling power reported. To enhance the reliability of the coating, BaSO$_4$-acrylic paint is developed with 60% volume concentration. The high filler concentration and broad particle size distribution help reach 98.1% solar reflectance and 0.95 sky window emissivity. During field tests, the BaSO$_4$ paint yields similarly high cooling power while providing great reliability, convenient paint form, ease of use and the compatibility with commercial paints.

**METHODS**

*BaSO$_4$ Nanoparticle Film and BaSO$_4$-Acrylic Paint Fabrication*

A BaSO$_4$ particle film of 150 μm thickness was fabricated on a silicon wafer. 400 nm BaSO$_4$ particles, deionized water and ethanol were mixed with a mass ratio of 2:1:1 and coated on the substrate until fully dried. The average particle size was chosen as 400 nm to reflect both visible and near infrared range of the solar irradiation. To fabricate the $BaSO_4$-acrylic nanocomposite paint, Dimethylformamide and $BaSO_4$ nanoparticles (400 nm diameter, US Research Nanomaterials) were mixed and ultra-sonicated for 15 minutes with a Fisherbrand Model 505 Sonic Dismembrator. The mixture was degassed to remove air bubbles introduced during the ultra-sonication process. Acrylic (Elvacite 2028 from Lucite International) was then slowly added and mixed until fully dissolved. The mixture was poured into a mold to be left fully dried overnight, resulting in a free-standing layer with a thickness of 400 μm to eliminate the substrate effect on the overall coating performance. A series of thinner coatings of the $BaSO_4$ film and paint were also prepared with film applicators (BYK Film Applicator) to study the effect of coating thickness. The dry film thickness was measured with a coordinate measuring machine (Brown&Sharp MicroXcel PFX).

*Spectral Emissivity Characterization*

The solar reflectance from 250 nm to 2.5 μm was measured with a Perkin Elmer Lambda 950 UV-Vis-NIR spectrometer with an integrating sphere. A certified Spectralon diffuse reflectance standard was used, and the solar reflectance was calculated based on the AM 1.5 solar spectrum.[38] The uncertainty of the solar reflectance was 0.5%, based on the measurement of five samples. The emissivity from 2.5 μm to 20 μm was characterized with a Nicolet iS50 FTIR with a PIKE Technology integrating sphere. The sky window emissivity was calculated based on the atmospheric transparent window of air mass 1.5 and water column 1.0 mm (IR Transmission Spectra, Gemini Observatory). The uncertainty of 0.02 was estimated based on the PIKE Technologies Mid-IR diffuse reflectance standard.

*Field Test*

There are two field testing setups made for cooling performance characterization on a building roof, as shown in Figure 5. Similar characterization setups were used in previous studies.[28–30] The setups were created from a block of white Styrofoam for thermal insulation and enclosed by silver mylar to reflect solar irradiation. A pyranometer (Apogee SP-510) was adopted to measure solar irradiation. The temperature measurement setup (Figure 5a) monitored the sample and ambient temperatures with T-type thermocouples. A thin layer of low-density polyethylene (LDPE) film functioned as a transparent cover against forced convection. The cooling power characterization setup (Figure 5b) included a feedback heater to characterize the cooling power by heating the samples to the ambient temperature, which minimized heat conduction and convection. Transparent side cover was added to reduce forced convection. Photo of both setups are shown in Figure 5c. They were placed on a high-rise table to eliminate heating from the ground.

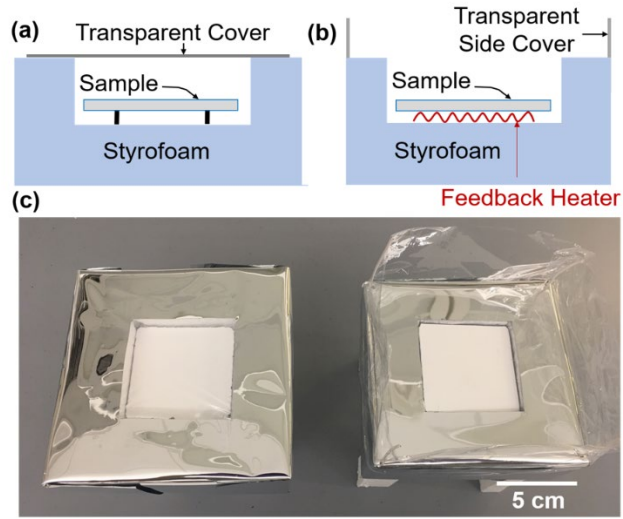

**Figure 5.** Field test setups for cooling performance characterization. (a) The samples were suspended in the Styrofoam cavity to minimize thermal conduction and forced convection. The sample and ambient temperatures were monitored during the temperature measurement. (b) To directly measure the cooling power of the sample, a feedback heater was included to heat the sample to the ambient temperature. The measured power consumption of the heater was the cooling power of the sample. (c) Photos of the field test setups. Both setups were located on a high-rise table to avoid ground heating effect.

*Monte Carlo Simulation*

The Monte Carlo simulations were performed according to the modified Lorentz-Mie theory [27] and a correction [33] of dependent scattering effect due to high concentrations. Photon packet is released at the top of the nanocomposite. The photon starts with a weight of unity and a normal direction to the air-composite interface. If the photon propagates to the bottom interface and makes it through, we consider that weight as transmitted. If the photon travels to the top surface and goes back to the air space above the medium, it is considered reflected. A total of 500,000 photons were used in each simulation, covering 226 wavelengths from 250 nm to 20 μm.

**ASSOCIATED CONTENT**

The supporting information includes Supplementary Note 1, 2, 3, Figure S1, S2 and Video S1.
Supplementary Note 1: Substrate-dependence of the $BaSO_4$ film samples

Supplementary Note 2: Energy balance model for the cooling power characterization

Supplementary Note 3: Cost analysis of the cooling paint

Figure S1. The effect of substrate on the sub-ambient sample temperature of the $BaSO_4$ film

Figure S2. The theoretical cooling power compared with experimental measurements of the $BaSO_4$ film and $BaSO_4$ paint

Video S1. Brushing, drying and water running test of the $BaSO_4$ paint

**AUTHOR INFORMATION**

**Corresponding Author**

Xiulin Ruan, Email: ruan@purdue.edu

URL: https://engineering.purdue.edu/NANOENERGY/

**Author Contributions**

Conceptualization, X.R.; Methodology, X.R., X.L. and J.P.; Investigation, X.L., J.P. and P.Y.; Writing - Original Draft, X.R. and X.L.; Writing - Review & Editing, X.R., X.L., J.P. and P.Y.; Funding Acquisition, X.R.; Resources, X.R.; Supervision, X.R. The manuscript was written through contributions of all authors. All authors have given approval to the final version of the manuscript.

**Notes**

The authors declare the following financial interests/personal relationships which may be considered as potential competing interests: X.R., X.L. and J.P. are the inventors of an international patent (PCT/US2019/054566) on the basis of the work described here.

ACKNOWLEDGMENT

The authors thank Dr. Mian Wang, Jacob Faulkner, Xuan Li, Nathan Fruehe, Daniel Gallagher and Professor Zhi Zhou at Purdue University for their help on sample fabrication and characterization. This research was supported by the Cooling Technologies Research Center at Purdue University and the Air Force Office of Scientific Research through the Defense University Research Instrumentation Program (Grant No. FA9550-17-1-0368).

**Supplementary Note 1: Substrate-dependence of the BaSO₄ film samples**

The 150 μm BaSO₄ film was coated on a silicon substrate, which only functioned as a supporting substrate, not to enhance the solar reflectance, nor to emit in the sky window. The substrate effect on the sub-ambient sample temperature is shown in Figure S1, where the different substrates had a negligible effect on the nighttime cooling performance of the 150 μm BaSO₄ film.

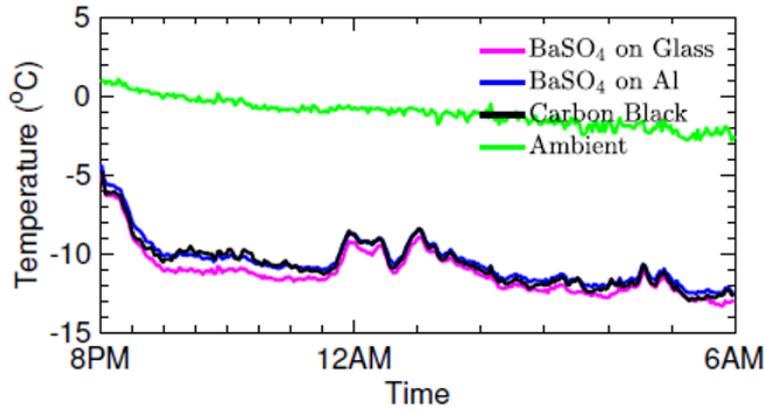

**Figure S1.** The effect of substrate on the sub-ambient sample temperature of the BaSO₄ film. The nighttime cooling performance of 150 μm BaSO₄ film on different substrates was measured and compared to a carbon black sample. All samples showed similar cooling performance, indicating the sky window emissivity was not affected by the substrate of the BaSO₄ film.

**Supplementary Note 2: Energy balance model for the cooling power characterization**

Using the temperature profiles in the direct cooling power characterization, we can analyze the cooling power and compare to the measured cooling power for validation. The net cooling power $q''_{cooling}$ according to the energy balance model is

$$q''_{cooling} = \frac{mC_p}{A}\frac{dT}{dt} - \alpha G + q''_{radiation}(T)$$

where $A$ is the surface area of the sample, $\frac{mC_p}{A}\frac{dT}{dt}$ accounts for the transient heat transfer due to the thermal mass of the sample and the heater, $\alpha$ stands for the solar absorption of the sample, $G$ represents the solar irradiation, $q''_{radiation}(T)$ is the emissive power through the sky window, and $q''_{cooling}$ is the net cooling power. $q''_{radiation}(T)$ term is mostly contributed by the thermal emission through the sky window, while the radiative exchange between the sample and the ambient is negligible. Because the sky window emission is highly dependent on the weather condition, $q''_{radiation}(T)$ is first calibrated with the nighttime cooling power measurement, which were 106 W/m² and 113 W/m² at 15°C during the field tests for BaSO$_4$ film and BaSO$_4$ paint, respectively. Using the experimentally measured $\alpha$, $G$, $T$, the theoretical net cooling power was modeled and compared with measured results in Figure S2. The model results agree reasonably well with the experimental measured cooling power, validating our field test results.

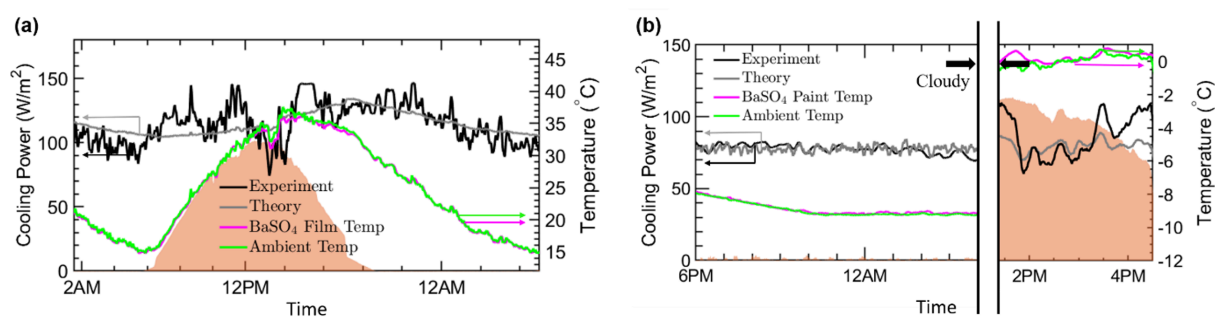

**Figure S2.** The theoretical cooling power compared with experimental measurements of the (a) BaSO$_4$ film and (b) BaSO$_4$ paint. The sky window emissive power was calibrated based on nighttime cooling power first, and the net cooling power was then estimated based on the experimental measured temperature profiles, solar irradiation and solar reflectance.

**Supplementary Note 3: Cost analysis of the cooling paint**

BaSO$_4$ is available in the natural mineral barite, and widely used in different industrial fields as radiocontrast agent, paper brightener and main components in cosmetic products [3]. BaSO$_4$ powders cost only $0.44 per kilogram [4], which is about half the price of TiO$_2$ powders ($1 per kilogram) [5]. Assuming a 30° roof angle and 300 μm paint thickness, the cost of the BaSO$_4$ fillers reaches around $100 for a 150 m² house. With a similar fabrication process to the commercial paints, the BaSO$_4$ paint will reach a comparable price as the commercial white paint.

In the literature, a comprehensive test was performed with commercial white roofing materials (85% solar reflectance) during one-month period in the summer, and concluded the energy saving was 40 to 75 Wh/m²/day [6]. The BaSO$_4$ cooling paint achieved a higher solar reflectance of 96.8% to 98.1%, leading to a total energy saving over 80 Wh/m$^2$/day, assuming daily solar energy as 5000 Wh/m$^2$/day and the AC stock average efficiency as 15. With the electricity cost about $0.1 per kWh, the monthly cooling cost saving can be $36 for a moderate 150 m$^2$ house.